\documentclass[twocolumn,prl]{revtex4}
\usepackage{graphicx}
\begin{document}
\title{Entanglement entropy and the Fermi surface}
\author{Brian Swingle}
\email{bswingle@mit.edu}
\affiliation{Department of Physics, Massachusetts Institute of Technology, Cambridge, MA 02139}
\begin{abstract}
Free fermions with a finite Fermi surface are known to exhibit an anomalously large entanglement entropy.  The leading contribution to the entanglement entropy of a region of linear size $L$ in $d$ spatial dimensions is $S\sim L^{d-1} \log{L}$, a result that should be contrasted with the usual boundary law $S \sim L^{d-1}$.  This term depends only on the geometry of the Fermi surface and on the boundary of the region in question.  I give an intuitive account of this anomalous scaling based on a low energy description of the Fermi surface as a collection of one dimensional gapless modes.  Using this picture, I predict a violation of the boundary law in a number of other strongly correlated systems.
\end{abstract}
\maketitle
\textit{Introduction}.---Currently ideas from quantum information theory are stimulating major developments in quantum many body theory and vice versa.  One of the major conceptual tools that has been receiving attention from both communities is entanglement entropy.  To define entanglement entropy the system must first be partitioned into two subsystems.  The von Neumann entropy $S = -\text{Tr}(\rho \ln{\rho})$ of the reduced density matrix of a subsystem is the entanglement entropy of the subsystem.  If the quantum state of the system factorizes over the two subsystems then the entanglement entropy is zero.  Also, if the whole system is in a pure state then the entanglement entropy is independent of which of the two subsystems is chosen.

Bosonic systems frequently satisfy an area or boundary law: the entanglement entropy is proportional to the size of the boundary of the subsystem \cite{arealaw1}.  This boundary law can be interpreted as arising from short range entanglement.  The boundary law is violated for conformal theories in one spatial dimension, and it is replaced by a logarithmic dependence $S \sim \log{L}$ on region size $L$ (boundary law behavior would correspond to a constant independent of region size) \cite{geo_ent,eeqft}.  The coefficient of this term is universal and proportional to the central charge of the conformal field theory. However, conformal field theories in higher dimensions so far satisfy a strict boundary law.

Interestingly, free fermions violate the boundary law even in more than one dimension \cite{fermion1}.  A sharp $d-1$ dimensional Fermi surface coincides with an extra logarithmic scaling of the entanglement entropy \cite{fermion2}.  Gapped fermionic systems and gapless Fermi systems with a higher codimension Fermi surface continue to obey a boundary law \cite{fermion3,fermion5}.  A detailed formula for the entanglement entropy of fermions with a sharp Fermi surface was proposed and found to be related to a conjecture of Widom important in signal processing \cite{fermion2}.  In this Letter I give an intuitive picture of the anomalous entanglement entropy of free fermions with a Fermi surface in agreement with the Widom formula.  The intuition is based on viewing free fermions as equivalent to a Fermi surface worth of $1+1$ dimensional chiral modes.  I also discuss other possibilities for observing violations of the boundary law.

\textit{Free fermions and chiral modes}.---I consider spinless fermions hopping on a square lattice in two dimensions.  The lattice provides an ultraviolet regulator for the theory.  It is possible to add spin and consider more generic lattices in different dimensions without any serious difficulties.  The physics is completely specified by giving the fermion dispersion relation $\epsilon_k$ as a function of the pseudomomentum $k$ lying in the 1st Brillouin zone.  In particular, the ground state is a Fermi sea where all states with energy less than the chemical potential $\mu$ are filled.  For fermions at half filling there is one fermion per two lattice sites and hence the Fermi sea will occupy half the Brillouin zone.  Generic filling fractions and dispersion relations lead to a Fermi surface with codimension $1$.  This means that the surface separating occupied and unoccupied regions in momentum space is a $d-1$ dimensional subspace of the the $d$ dimensional Brillouin zone.  The generic presence of a finite density of states at the Fermi surface is responsible for many of the unusual properties of the Fermi gas.

As an example of such an unusual property, the Fermi gas has a heat capacity linear in $T$ in any dimension $d$.  This result may be contrasted with the strongly dimension dependent result for superfluid bosons $C \sim T^d$.  My interest here is in the anomalous entanglement entropy of the Fermi gas.  Many bosonic and gapped fermionic systems in $d>1$ spatial dimensions have an entanglement entropy scaling as $L^{d-1}$ for a region of size $L$, but free fermions with a codimension $1$ Fermi surface have the anomalous scaling $L^{d-1} \ln{L}$.  Remarkably, a precise formula for the logarithmic term in the entropy was conjectured and verified numerically \cite{fermion2,fermion3,fermion5}.  This formula reads
\begin{equation}
S = \frac{L^{d-1}}{(2\pi)^{d-1}} \frac{\log{L}}{12}\int \int |n_x \cdot n_k | dA_x \, dA_k,
\end{equation}
where $n_x$ and $n_k$ are unit normals for the real space boundary and the Fermi surface respectively.  The integrals are over a scaled version of the real space boundary (hence the overall $L^{d-1}$ factor) and the Fermi surface, and the whole expression is written in units where the volume of the Fermi sea is one. I would like to understand how this anomaly arises from the presence of the Fermi surface.

Because of the finite density of states at the Fermi surface it is useful to reinterpret the low energy modes in terms of a large number of decoupled ``radial" excitations.  The word radial is used because the Fermi velocity $v_k = \nabla_k \epsilon_k$ is normal to the Fermi surface.  Thus I can assign to each point on the Fermi surface a radial fermionic mode with approximately linear dispersion and velocity given by the local Fermi velocity.  These modes are effectively relativistic and $1+1$ dimensional traveling in the local radial direction as a function of time.  They are also chiral because the direction of propagation is fixed by the local Fermi velocity.  Modes traveling in the opposite direction are typically on the other side of the Fermi surface.  Chirality is equivalent to the statement that there are no holes above the Fermi surface.  This kind of patching procedure is the first step towards a higher dimensional analog of bosonization, but I will not need anything more than the heuristic picture of many chiral excitations.  This point of view is also visible in various renormalization group treatments of Fermi liquid systems \cite{fermion_rg3,fermion_rg2,fermion_rg1}.

\textit{Entanglement entropy}.---The presence of a large number of chiral one-dimensional excitations at low energy strongly suggests a violation of the boundary law.  To make this precise I need some way to count the effective number of such chiral modes.  This counting can be performed using some intuition from the study of entanglement entropy, namely that the imaginary partitions introduced to compute the entanglement entropy behave very much like real physical boundaries.  Tracing out degrees of freedom outside a region of characteristic size $L$ should coarse grain the Fermi surface into patches of typical size $1/L^{d-1}$ exactly as in a finite size system.  Each patch contributes a factor of $\log{L}$ to the entanglement entropy because of the presence of a gapless one dimensional mode.  The total contribution from the Fermi surface should therefore scale as $L^{d-1} \log{L}$ with the Fermi momentum making up the units where necessary.  This simple argument is made more precise below, but it already captures the basic intuition behind the violation of the boundary law.

\begin{figure}
\includegraphics[width=.5\textwidth]{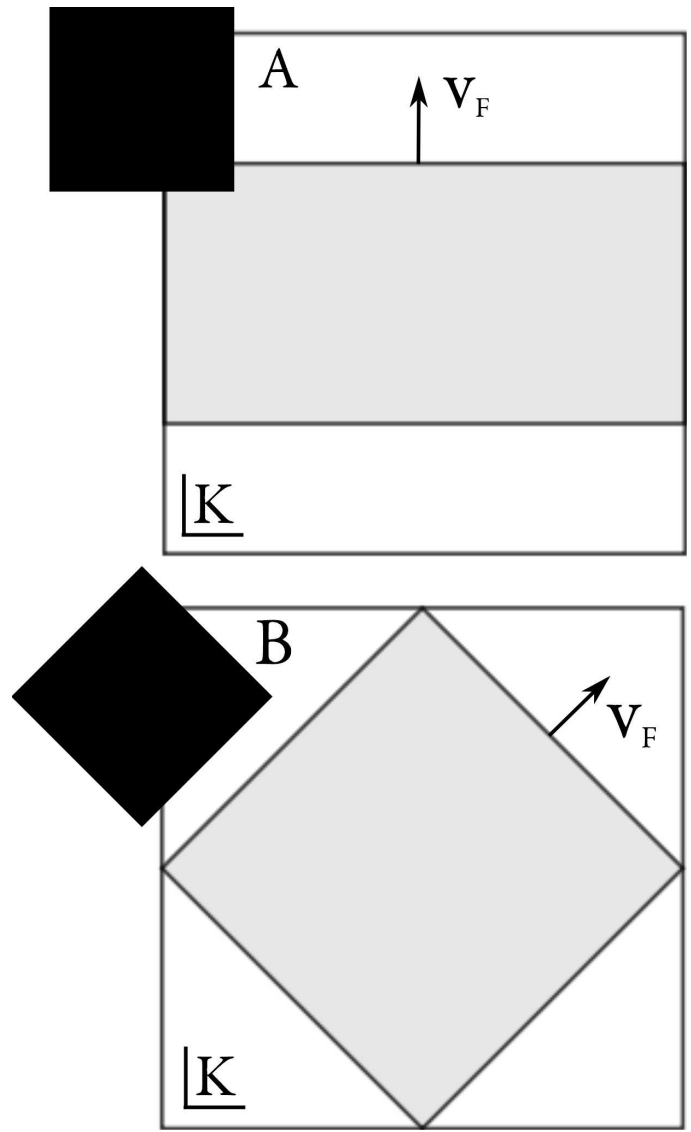}
\label{fig_2}
\caption{A sketch of the two model systems considered.  Box A shows the Fermi sea (in gray) of the strongly anisotropic model.  The relative orientation of the Fermi velocity and the chosen real space region (in black) is shown.  Box B shows the Fermi sea (in gray) at half filling for a fermion hopping on a square lattice. Again, the relative orientation of the Fermi velocity and the chosen real space region (in black) is shown.}
\end{figure}

Consider the case of half filling with dispersion $\epsilon_k = -2 \epsilon_0 \cos{k_y a}$ arising from a strongly anisotropic lattice model where fermions are unable to hop in the $x$ direction.  The Fermi surface is nested and has two disconnected components given by the lines $k_y = \pi/2a$ and $k_y = -\pi/2a$ where $a$ is the lattice spacing.  Let the subsystem of interest be a box-like region of dimensions $L\times L$ aligned with the $x$ and $y$ axes.  The mode density on the Fermi surface is $L/2\pi$ and the length of the Fermi surface is $2\pi/a + 2\pi/a = 4\pi/a$ for a total of $2L/a$ modes.  Each mode is chiral with left and right central charges given by $c_L = 1$ and $c_R = 0$.  Left and right are defined locally on each patch relative to Fermi velocity, but the important point is only that one of the central charges is one while the other is zero.  The entanglement entropy of a one dimensional conformal field theory on an interval of length $L$ is $\frac{c_L + c_R}{6} \log{(L/ \epsilon)}$ \cite{geo_ent,eeqft}.  Putting everything together I find a total entropy of $(L/a)\log{L}/3$.  In order to compare with previous work I use units where the volume of the Fermi sea is one.  This requirement is $4\pi^2/(2 a^2) = 1$ (the $2$ is for half filling) giving $a = \sqrt{2} \pi$.  I find a total entropy of $S = \frac{1}{3\sqrt{2}\pi} L \log{L}$ in agreement with the Widom formula.

Another simple situation is the case of equal hopping with dispersion $\epsilon_k = -2 \epsilon_0 \cos{k_x a} - 2 \epsilon_0 \cos{k_y a}$.  At half filling the Fermi surface consists of a square rotated by $45$ degrees occupying half the Brillouin zone.  To again make things simple consider an $L\times L$ region rotated by $45$ degrees so that it is oriented identically to the Fermi surface.  Similar mode counting arguments now give $2 \sqrt{2} L /a$ modes.  Using units where $a = \sqrt{2} \pi$, I find a total entropy of $S = \frac{1}{3 \pi} L \log{L}$ again in agreement with the Widom formula.


After these examples an understanding of the general Widom formula can be obtained by breaking the real space boundary into small segments.  I focus on the two dimensional case to make the notation as simple as possible.  Consider a segment $\Delta A_x$ of the real space boundary and a segment $\Delta A_k$ of the Fermi surface.  With a mode density of $\Delta A_x/2\pi$ the patch $\Delta A_k$ contributes
\begin{equation}
\frac{\Delta A_x \Delta A_k |n_x \cdot n_k |}{2 \pi}
\end{equation}
modes.  The flux factor $|n_x \cdot n_k |$ counts the number of modes perpendicular to the real space boundary.  In $d$ dimensions the above formula is modified by replacing $2\pi$ with $(2 \pi)^{d-1}$ since $\Delta S_x$ is now a general $d-1$ dimensional surface element.  Each of these modes is chiral and contributes
\begin{equation}
\Delta S \sim \frac{c_L + c_R}{6} \log{L_x} = \frac{1}{6} \log{L}
\end{equation}
to the entanglement entropy on an interval of length $L$.  Note that the precise choice of linear size $L$ in the logarithm is not critical as differences can be absorbed into the non-universal boundary law piece of the entanglement entropy.

\begin{figure}
\includegraphics[width=.5\textwidth]{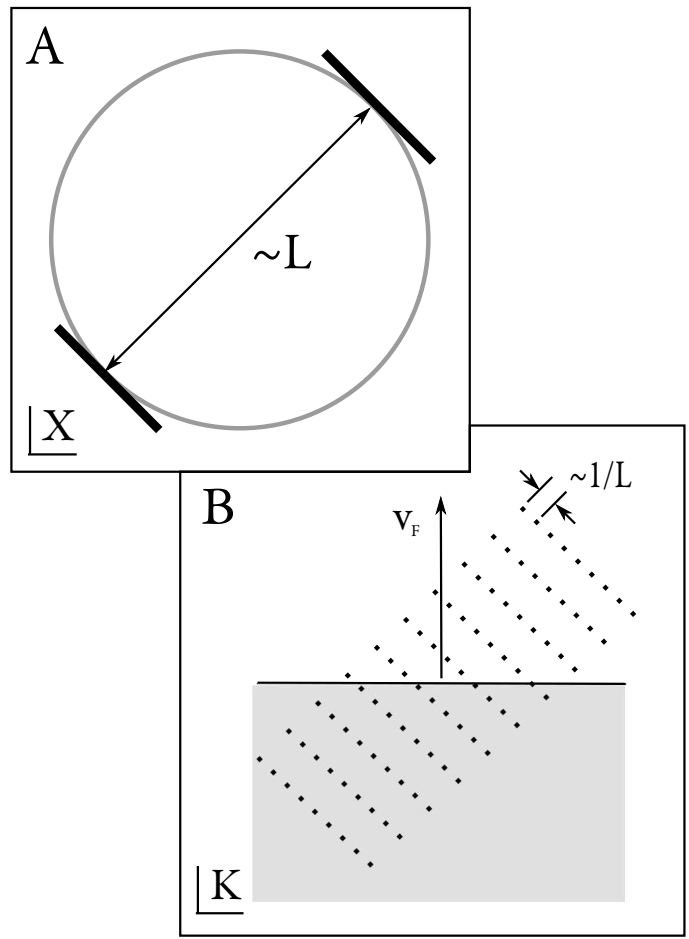}
\label{fig_2}
\caption{Box A shows a spherical real space region of linear size $L$ along with a chosen real space patch on the boundary.  The size of the patch (bold black bar) has been exaggerated for clarity.  Box B shows a local piece of the Fermi sea with filled states in gray.  The dots are an exaggerated representation of the effective mode quantization coming from the real space patch.  The relative orientation of the Fermi velocity $\propto n_k$ and the real space patch normal $n_x$ is the origin of the ``flux factor" $|n_x \cdot n_k |$ counting the effective number of modes that propagate perpendicular to the real space patch normal $n_x$.  Here the angle between $n_x$ and $n_k$ is $\pi / 4$.}
\end{figure}

The total entanglement entropy is given by integrating the contributions from all patches of the real space boundary and the Fermi surface.  This result must multiplied by an additional factor of $1/2$ because intervals are double counted in the integration.  Equivalently, each end of a one dimensional interval can be thought of as contributing $\frac{c_L + c_R}{12} \log{(L/\epsilon)}$ to the entropy \cite{eeqft}.  The full result is
\begin{equation}
S \sim \frac{\log{L}}{6} \frac{1}{2} \int \int \frac{dA_x dA_k}{(2\pi)^{d-1}} |n_x \cdot n_k|,
\end{equation}
and this is the Widom formula before rescaling the real space integral.

\textit{Discussion}.---I have shown in some simple cases that the entanglement entropy for free fermions can be obtained using counting arguments and intuition from one dimensional systems.  I also argued that an explicit formula based on the Widom conjecture is the correct generalization to arbitrary Fermi surface shape and region geometry.  These arguments naturally reproduce the fully one dimensional case exactly.  It is remarkable that the central charges of the chiral fermion are just right to make the Widom formula true.  Nevertheless, the value of the present point of view is not that it provides a rigorous derivation but that it gives intuition to search for other violations of the boundary law.

The theory described here predicts a violation of the boundary law in any system with a large number of gapless $1+1$d modes.  Fermi liquids certainly qualify.  The recently proposed d-wave Bose liquid phase should also violate the boundary law \cite{dbl}.  Similarly, non-Fermi liquids with a sharp Fermi surface but no Landau quasiparticle should violate the boundary law \cite{crit_fs}.  Gauge/gravity duality has recently provided an interesting example of such a strongly correlated non-Fermi liquid phase \cite{nfl1}.  Frustrated quantum magnets where the low energy description is in terms of deconfined spinons with a Fermi surface should also violate the boundary law.  In fact, this violation may be a useful numerical test for the existence of such a spinon phase.  This is because the logarithmic correction to the boundary law is a low energy phenomenon that grows faster with region size than the non-universal boundary law term.

Finally, I mention a few other implications of the theory.  On the numerical side, the systems mentioned above should be difficult to simulate using tensor network approaches \cite{mera,peps,terg}.  Such schemes are tailored to boundary law behavior in the entanglement entropy, and no natural alternative is available for systems that violate the boundary law in more than one dimension.  On the other hand, the violation of the boundary law could be a useful numerical signature of such phases using exact diagonalization or other methods that are not primarily limited by entanglement.

Another implication concerns the appearance of the boundary law violating term across a phase transition from a gapped phase.  The gapped phase near the phase transition can be viewed as a Fermi surface worth of gapped one dimensional modes.  The simplest example would be a superconducting to Fermi liquid phase transition where $\xi$ is related to the inverse of the superconducting gap.  Replacing $L$ by the correlation length $\xi$, the gapped modes contribute an entropy $L^{d-1} \log{\xi}$ and obey the boundary law.  As the quantum critical point is approached, the diverging boundary law term in the entropy is cutoff when $\xi$ exceeds $L$.  The result is a boundary law violating term of the form $L^{d-1} \log{L}$ as before.

\textit{Acknowledgements}.---I would like to thank T. Grover, M. Barkeshli, and Senthil for helpful discussions.  I would like to thank X.-G. Wen for support and encouragement during this work.  This research was supported in part by Perimeter Institute for Theoretical Physics.

\bibliography{fermions_ee}
\end{document}